\documentclass[twocolumn,pra,aps]{revtex4}
\usepackage{graphicx}

\begin{document}

\title{On Bell, Suarez-Scarani, and Leggett experiments:\\
Reply to a comment, and proposal for a new experiment}

\author{Antoine Suarez}
\address{Center for Quantum Philosophy, P.O. Box 304, CH-8044 Zurich, Switzerland\\
suarez@leman.ch, www.quantumphil.org}

\date{September 20, 2008}

\begin{abstract} It is shown that the before-before (or Suarez-Scarani) experiment refutes hidden variable models with a deterministic (``realistic'') nonlocal part, and the experimental violation of Leggett-type inequalities models with a random nonlocal part and a biased random local one. Therefore the claim that Gr\"{o}blacher et al. present ``an experimental test of nonlocal realism'' \cite{gro} is misleading, and Marek \.{Z}ukowski's comment \cite{zu} misses the point. A new experiment is proposed.\\

\end{abstract}


\maketitle

In a recent comment \cite{zu} Marek \.{Z}ukowski, co-author of Reference \cite{gro}, claims  that I wrongly present in Reference \cite{as08} the assumptions behind the Leggett's inequalities, and their modified form used by Gr\"{o}blacher et al. \cite{gro} for an experimental falsification of a certain class of non-local hidden variable models. Nevertheless, \.{Z}ukowski's  writes that his comment ``is not aimed at a detailed discussion of the arguments given by Suarez'' \cite{zu}.

By omitting details \.{Z}ukowski's comment misses the point. What I state is this:

The authors of \cite{gro} do not only say that their experiment violates the Leggett's inequalities, but also claim that it is ``an experimental test of nonlocal realism''. More specifically, they claim that their result ``excludes for the first time a broad class of non-local hidden-variable theories'' \cite{gro}.

Against this claim I argue:\\

1) The before-before experiment \cite{asvs97, as00.1, szsg} was first in excluding a class of nonlocal hidden variable theories, and thus ''nonlocal realism''.\\

2) Gr\"{o}blacher et al. did not test ``nonlocal realism''.

\begin {center}
\emph{Regarding 1)}
\end {center}

Bell type experiments refuted local hidden variable models. However, by adding nonlocal hidden variables it is still possible to save determinism (see, e.g., \cite{jb}). This is the case in the Suarez-Scarani model \cite{asvs97,as00.1,vang}. This model uses moving measuring devices, and thereby different relativistic timings. It assumes time-ordered nonlocal dependencies (nonlocal determinism) for certain timings, and only local hidden variables (local determinism) for the before-before timing. This leads to conflict with the timing independence of quantum mechanics. The before-before experiment falsified the nonlocal deterministic model of Suarez-Scarani, and confirmed the quantum prediction \cite{as00.1,szsg}.

Thus, the Suarez-Scarani experiment was first in excluding nonlocal hidden variable models. It showed that, to borrow a phrase from \cite{gro}, ``giving up the concept of locality is not sufficient to be consistent with quantum experiments''.

``Realism'' as defined in Reference \cite{gro} has the meaning of determinism. I do not say that Gr\"{o}blacher et al. support ``gender asymmetry'' \cite{zu}, but nonlocal determinism \cite{bohm}. They assume that Alice's outcome is nonlocally \emph{predetermined} by Bob's one, or Bob's outcome by Alice's one. The ``explicit toy non-local model'' \cite{zu} Gr\"{o}blacher et al. propose clearly shows that they have  nonlocal deterministic models in mind (toys are usually good indicators of cognitive structures).

Thus, the model described in Reference \cite{gro} can be considered refuted by the before-before experiment (unless one postulates a single preferred frame what, on the one hand, is not the case in \cite{gro}, and on the other hand, bears severe oddities \cite{as08}).

\begin {center}
\emph{Regarding 2)}
\end {center}

In Leggett models the hidden variables have always a local and a nonlocal part \cite{cb,core}, independently of any timing.

As soon as one assumes a \emph{deterministic nonlocal} part, the Suarez-Scarani experiment becomes obviously relevant for Leggett's models and refutes them.

Thus, the specific aim of experiments testing Leggett-type inequalities is to test models exhibiting \emph{nonlocal randomness}, and non-trivial local parts, i.e., outcomes that depend on biased random local variables \cite{cb,core}.

This means that Gr\"{o}blacher et al., in spite of assuming ``nonlocal realism'', in fact did not test this assumption.

Certainly, their experiment \cite{gro} (as far as its implementation is correct) would also rule out a model without determinism in the nonlocal part, and so is useful in addition to the before-before experiment when interpreted correctly \cite{rc}. And in any case, has the merit of priority as a proposal to test Leggett inequalities.

\begin {center}
\emph{Conclusion and proposal for a new experiment}
\end {center}

The Suarez-Scarani (before-before) experiment excludes \emph{time-order} or determinism in the nonlocal part. That is, the quantum correlations come from outside spacetime through free choices in Nature (God plays dice) \cite{az05}.

Legget experiments demonstrate that Nature refuses even to \emph{mimic} certain deterministic (``realistic'') features by means of biased random local variables (God plays \emph{fair} dice).

Putting together the results of both types of experiments one can conclude that Nature is not less random than predicted by quantum mechanics. \cite{bra}

Nevertheless, to date the results supporting this conclusion have been gathered in separated experiments. It would be suitable to refute both determinism and biased randomness by one and the same experiment. I think this may be nicely done by a before-before version of the experiment described by Colbeck and Renner \cite{core}. The new experiment would basically consist in demonstrating \emph{firstly}, that Nature exhibits correlations originating from pure nonlocal links and \emph{secondly}, these links are not time-ordered. Work exploring this possibility is in progress. Such an experiment would definitely contribute to a better understanding of the quantum.\\

\noindent\emph{Request}: Though the Suarez-Scarani experiment was first in testing nonlocal determinism and excluding nonlocal hidden variable models, the experiment was not quoted in Reference \cite{gro}. Thereby Gr\"{o}blacher et al. overlooked relevant work and advanced a misleading interpretation of their own results. I think \emph{Nature}'s general audience deserves to be informed about this state of affairs, and kindly request Anton Zeilinger and the Editor to agree in publishing a clarifying comment.\\

\noindent\emph{Acknowledgments}: I am grateful to Marek \.{Z}ukowski for his stimulating comment, and  Cyril Branciard, \v{C}aslav Brukner, Roger Colbeck, Bernard d'Espagnat,  Nicolas Gisin, Valerio Scarani, Hugo Zbinden for discussions on the subject.

\end{document}